# Distributed simulation of polychronous and plastic spiking neural networks: strong and weak scaling of a representative mini-application benchmark executed on a small-scale commodity cluster

Pier Stanislao Paolucci[1*], Roberto Ammendola[2], Andrea Biagioni[1], Ottorino Frezza[1], Francesca Lo Cicero[1], Alessandro Lonardo[1], Elena Pastorelli[1], Francesco Simula[1], Laura Tosoratto[1], Piero Vicini[1]

[1]INFN Roma "Sapienza", Italy

[2]INFN Roma "Tor Vergata", Italy

[*]Corresponding author: Pier Stanislao Paolucci, E-mail pier.paolucci@roma1.infn.it

## ABSTRACT

We introduce a natively distributed mini-application benchmark representative of plastic spiking neural network simulators. It can be used to measure performances of existing computing platforms and to drive the development of future parallel/distributed computing systems dedicated to the simulation of plastic spiking networks. The mini-application is designed to generate spiking behaviors and synaptic connectivity that do not change when the number of hardware processing nodes is varied, simplifying the quantitative study of scalability on commodity and custom architectures. Here, we present the strong and weak scaling and the profiling of the computational/communication components of the DPSNN-STDP benchmark (Distributed Simulation of Polychronous Spiking Neural Network with synaptic Spike-Timing Dependent Plasticity). In this first test, we used the benchmark to exercise a small-scale cluster of commodity processors (varying the number of used physical cores from 1 to 128). The cluster was interconnected through a commodity network. Bidimensional grids of columns composed of Izhikevich neurons projected synapses locally and toward first, second and third neighboring columns. The size of the simulated network varied from 6.6 Giga synapses down to 200 K synapses. The code demonstrated to be fast and scalable: 10 wall clock seconds were required to simulate one second of activity and plasticity (per Hertz of average firing rate) of a network composed by 3.2 G synapses running on 128 hardware cores clocked @ 2.4 GHz. The mini-application has been designed to be easily interfaced with standard and custom software and hardware communication interfaces. It has been designed from its foundation to be natively distributed and parallel, and should not pose major obstacles against distribution and parallelization on several platforms. The DPSNN-STDP mini-application benchmark is developed in the framework of the EURETILE FET FP7 European project, in cooperation with the CORTICON FET FP7 project.[1]

## 1. Introduction

Brain simulation is: 1- a scientific grand-challenge; 2- a source of requirements and architectural inspiration for future parallel/distributed computing systems, 3- a parallel/distributed coding challenge. The main focus of several neural network simulation projects is the search for a)-biological correctness; b)-flexibility in biological modeling; c)-scalability using commodity technology [e.g., NEURON (Carnevale, 2006, 2013); GENESIS (1988, 2013); NEST (Gewaltig, 2007);]. A second research line focuses more explicitly on computational challenges when running on commodity systems, with varying degrees of association to specific platforms echo-systems [e.g., Del Giudice, 2000; Modha, 2011; Izhikevich, 2008, Nageswaran, 2009]. An alternative research pathway is the development of specialized hardware, with varying degrees of flexibility allowed [e.g. SPINNAKER (Furber, 2012), SyNAPSE or BlueBrain projects]. Since 1984, the focus of our APE lab at INFN is the design and deployment of parallel/distributed architectures dedicated to numerical simulations (e.g. Avico et al., 1986; Paolucci, 1995). The present center of interest of APE lab is the development of custom interconnection networks (R Ammendola et al ., 2011). Indeed, the original purpose of the DPSNN-STDP project is the development of the simplest yet representative benchmark (i.e. a mini-application), to be used as a tool to characterize software and hardware architectures dedicated to neural simulations, and to drive the development of future generation simulation systems. Coded as a network of C++ processes, it is designed to be easily interfaced to both MPI and other (custom) Software/Hardware Communication Interfaces. The DPSNN-STDP mini-application benchmark has been designed to be natively distributed and parallel, and should not pose obstacles against distribution and parallelization on several competing platforms. It should capture major key features needed by large cortical simulations, should serve the community of developers of dedicated computing systems and should facilitate the tuning of commodity platforms. Moreover, the code demonstrated to be fast and scalable. For example, 115 wall clock seconds are required to simulate one second of activity and plasticity of a network composed by 1.6 G synapses (8 M neurons spiking at a mean firing rate of 22.5 Hz) on 128 hardware cores running @ 2.4 GHz. One of the explicit objectives is to maintain the code readable and its size at a minimum, to facilitate its usage as a benchmark.

---

[1] The EURETILE project is funded by the European Commission, through the Grant Agreement no. 247846, Call: FP7-ICT-2009-4 Objective FET-ICT-2009.8.1 Concurrent Tera-device Computing. See Paolucci et al, 2013. The CORTICONIC project is funded through the FET FP7 Grant Agreement no. 600806.

This document presents: 1- measures of the strong and weak scaling behavior of the January 2014 release of the DPSNN-STDP mini-application benchmark, run on a small scale cluster of commodity processors interconnected through a commodity network; 2- the profiling of the time-spent on main code components (computational and inter-process communication blocks); 3- how the execution time changes depending on the number of synapses projected by each neuron; 4- hints for scalability derived from the profiling of the execution, that we think could be generalized to several simulators. The "Methods" section of this document provides a compact description of the features of the current release of the DPSNN-STDP mini-application benchmark. The "Results" section reports the strong and weak scaling and code profiling measures produced on a small scale commodity cluster. We analyze the results and the hints for future developments in the "Discussion" and "Conclusions" section of the document.

During 2014, we will further enhance DPSNN-STDP: 1- to enable the description of more complex connectomes; 2- to support other neural/synaptic models; 3- to prepare it for a possible distribution to a larger community of users.

## 2. Methods

Here, we provide a compact summary about the internal structure of the DPSNN-STDP mini-application benchmark. A more complete technical report will be published later on, during 2014.

**Each process describes a cluster of neurons and their incoming synapses**

The full neural system is described by a network of C++ processes equipped with a message passing interface, agnostic of the specific message passing library. The full network is divided into clusters of neurons and their set of incoming synapses. The data structure that describes the synapse includes the information about the total transmission delay introduced by the axonal arborization that reaches it. The list of local synapses is further divided in sets according to the value of the axo-synaptic delay. Each C++ process describes and simulates a cluster of neurons and incoming synapses. The messages travelling between processes are sets of "axonal spikes" (i.e. they carry info about the identity of neurons that spiked and the original emission time of each spike). Axonal spikes are sent only toward those C++ processes where at least a target synapse exists for the axon. The knowledge of the original emission time of each spike and of the transmission delay introduced by each synapse allows for the management of synaptic STDP (Spike Timing Dependent Plasticity) (Song, 2000), which produces effects of Long Term Potentiation/Depression (LTP/LTD) of the synapses. This temporal information produces phenomena related to difference among delays along (chains of) individual synaptic arborizations (polychronism, see Izhikevich, 2006).

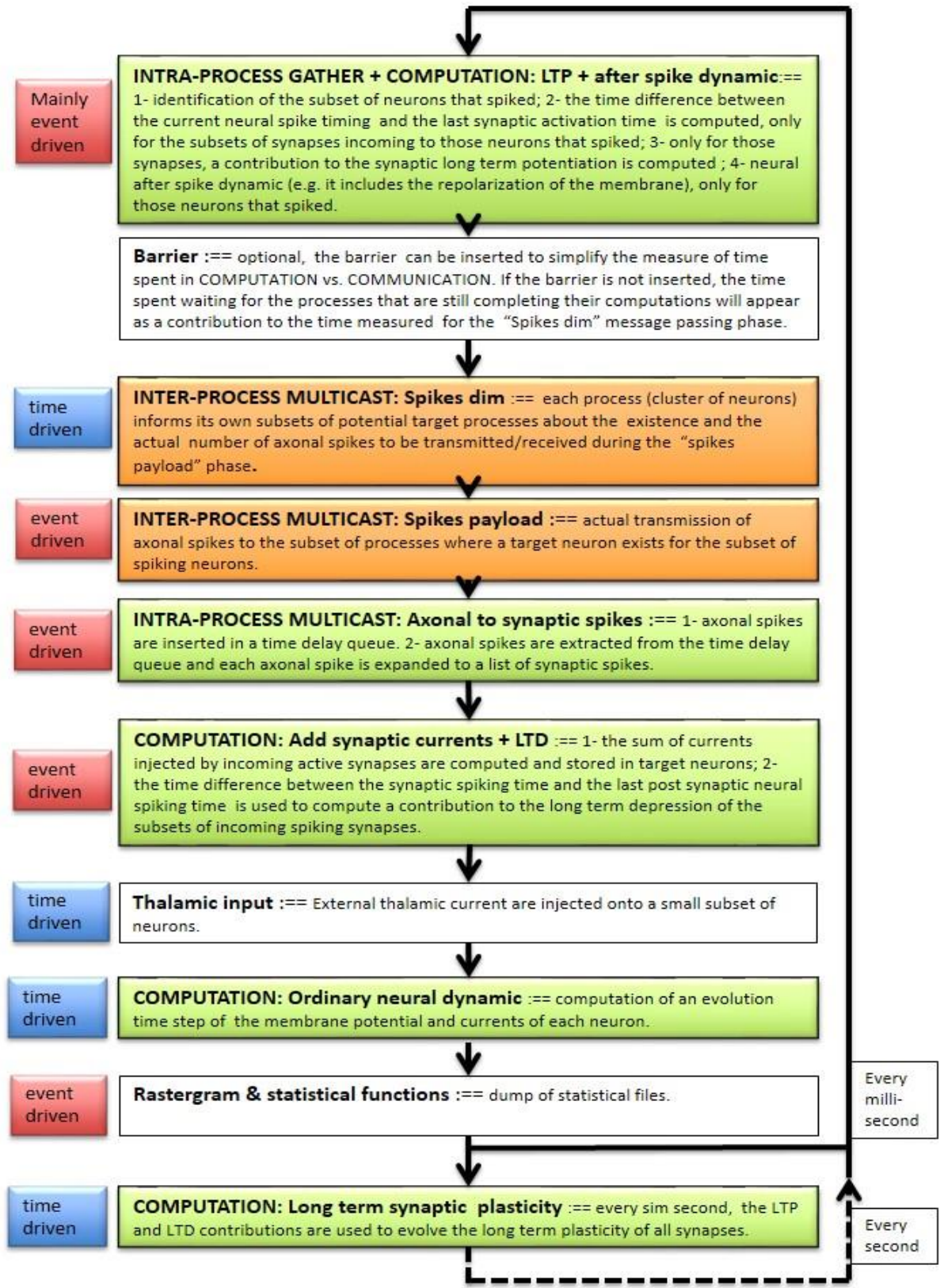

Figure 2-1. Cluster of neurons and incoming synapses are assigned to software process in a distributed simulation. Each process iterates over the blocks represented in this picture, that simulate the dynamic of neurons, the spiking and plasticity of synapses and the exchange of messages through axo-dendritic arborization. It is a flow of event-driven and time-driven computational and communication (inter-process and intra-process multicast) blocks. The measure of the (relative) execution times of the blocks guided the effort dedicated to the optimization of each block and can drive the development of dedicated hardware architecture.

### Mixed time-driven and event driven simulation

There are two phases in a DPSNN-STDP simulation: 1- the creation of the initial state and structure of the neural network; 2- followed by the simulation of the dynamic of neurons and synapses. The initial construction of the system (1) includes the creation of the network of axonal polychronous arborizations and synapses that interconnect the system.

We adopted a combined event-driven and time-driven approach (Morrison et al, 2005) for the simulation of the neural and synaptic dynamic:

- *Event-driven simulation*, for synaptic dynamics.
- *Time-driven simulation*, for neural dynamics.

The phase of simulation of the dynamic (2) can be further decomposed into an iteration over the following steps: 2.1- the subset of neurons that produced spikes during the previous time-driven simulation step induces an event-driven long term potentiation of their incoming synapses using the STDP scheme and perform other post-spiking activities; 2.2 - spikes are sent through axonal arborizations to the cluster of neurons where target synapses exist; 2.3 - axonal spikes delivered to a process are queued into a list, for usage during this time-step and a window of subsequent time steps; 2.4 - axonal spikes, classified according to their time of arrival, are delivered through local axonal arborizations to the subset of their target synapses of appropriate delay; 2.5 - synapses inject currents into their target neuron; 2.6 - target neurons induce an event-driven long term depression on the subset of synapses that injected currents; 2.7 - the full set of neurons perform one step of time evolution.

### Spiking neuron model

Hybrid models describe the continuous evolution of several state variables (including a "membrane voltage" and auxiliary "currents") and discrete events associated to the spiking event, i.e. special rules applied to (a subset of) the state variables. Well known are the Hodgkin-Huxley (HH) (Huxley, 1952), the leaky integrate-and-fire (LIF) and the Izhikevich (IZH) (Izhikevich, 2003). For this experiment we adopted the IZH model which is computationally efficient (13 – 26 operations per simulated ms per neuron), and yet capable of replicating the spiking behaviour of several neuron types (Izhikevich, 2004).

$$\begin{cases} if\ v(t) < v_{peak}\quad then\ \begin{cases} \dot{v} = 0.04\ v^2 + 5v + 140 - u + I \\ \dot{u} = a(bv - c) \end{cases} \\ if\ v(t) \geq v_{peak}\quad then\ \begin{cases} v(t + \Delta t) = c \\ u(t + \Delta t) = u(t) + d \end{cases} \end{cases}$$

where:
- v(t) represents the neural membrane potential. We say that when *v* reaches $v_{peak}$ a "neural spike" happened;
- I(t) is the potential change generated by the sum of all synapses incoming to the neuron. Incoming currents are present if spikes arrived form presynaptic neurons;
- u(t) represents a membrane recovery variable;
- a, b, c, d are four parameters, constant for each neuron kind, by varying them the same equation models several kind of known neural types.
- After a spike, the membrane potential and the recovery variable are reset using the c and d reset constants;

In this experiment we used a mix of 80% excitatory RS Izhikevich neurons (i.e.: a=0.02, b=0.2, c = -65.0 mV, d=8.0) and 20% inhibitory FS neurons (obtained by setting a=0.1, b=0.2, c = -65.0 mV, d=2.0). $v_{peak}$ was set at 30 mV.

**Synaptic update: spike-timing dependent plasticity**

Let us define t = $t_{post}$ − $t_{pre}$ − $d_{axon}$, the time difference between the post-synaptic spike time, and the time of arrival of a spike originated by a presynaptic neuron at an original emission time $t_{pre}$, that arrives at the target after an axonal delay $d_{axon}$. We implemented the following STDP rule, to compute the $\Delta W_{pre,post}$ change to the synaptic strength (Song et a., 2000). $A_+$, $A_-$, $\tau_+$, $\tau_-$, are parameters which permits to match the model on different types of neurons and biochemical contexts.

$$t = t_{post} - t_{pre} - d_{axon} \quad \begin{cases} if \ t \geq 0 & \Delta W_{pre,post} = A_+ e^{-\frac{t}{\tau_+}} \\ if \ t < 0 & \Delta W_{pre,post} = A_- e^{\frac{t}{\tau_-}} \end{cases}$$

The synapse is maximally potentiated if the delay introduced by the axon carries the signal to the target just before the post-synaptic spike (i.e. it is probably the cause of the spike). The synapse is maximally depressed if the signal arrives just late.

**Projection of synapses with individual delay**

In our polychronous networks each neuron $i = 1..N$ projects its set of forward synapses $j=1..M$, each one characterized by its individual delay $D^{i,j}$, plastic weight $W^{i,j}$ and target neuron $K^{i,j}$. In this set of tests, M was fixed to 200 for all neurons. Inhibitory neurons projected synapses only toward excitatory neurons located in the same column. Instead, excitatory neurons projected also to neighboring columns, as discussed in a following section. For this experiment, we assigned delays in the range between 1 and 20 ms. Inhibitory synapses were assigned with the minimum delay, while excitatory delays were assigned with a uniform distribution of delays. If a neuron fires at a time $t_i$, each forward synapse will have to inject a current $W^{i,j}$ at a synaptic specific time $t_i + D^{i,j}$. The current $W^{i,j}$ will be injected in the target neuron $k_j=K^{i,j}$, where it will add to the currents arriving in the same time step from other source neurons, to form the total external incoming current $I_k(t)$ which contributes to the dynamic of neuron $k_j$.

**Bidimensional arrays of neural columns and their distribution on the processes and processors**

In this experiment of strong and weak scaling behavior, we arranged the neurons in "columns", each one composed by one thousand neurons. Columns are then arranged in bidimensional grids (see Figure 2-1). Each excitatory neuron projected 76% of its synapses to neurons in its own column, 12% of its synapse toward neurons in first neighboring columns, 8% towards second neighbors and 4% toward neurons in third neighboring columns (see picture). Inhibitory neurons projected their synapses only towards excitatory neurons in the same column. We varied the total number of columns between one and 32768 (a 256x128 grid of neural columns, for a total of 32.7 million neuron). Each process can either host a fraction of a column (e.g. 1/8, ¼ …), a whole single column, or several columns (in the present version, up to 32 columns per process). When performing strong and weak scaling measures, periodic boundary conditions are used for columns at the boundary of the grids, to get more homogeneous spiking rates for different numbers of columns. For grids so small not to have enough distinct neighbors, the periodic boundary rule can end projecting more synapses on the same target column than expected for a large grid. Actually, in the case of a single column, all synapses are projected by the column to itself.

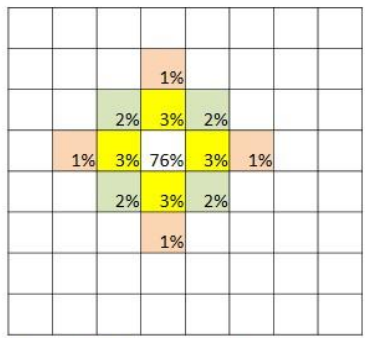
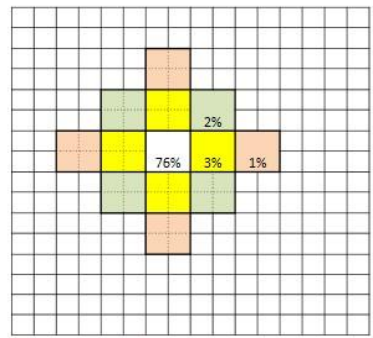
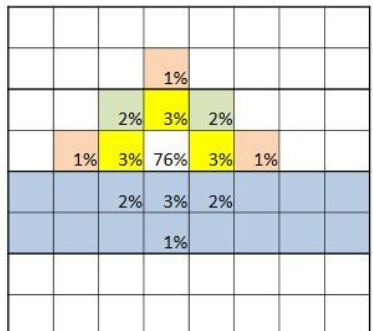

Figure 2-2. Example of distribution of an identical problem over a varying number of software processes and computational cores. In this figure, the grid to be distributed is composed by 8x8=64 neural columns. The DPSNN-STDP simulator produces the same external stimulus, synaptic structure and spiking activity on all distributions. We varied the number of columns (from 1 to 32768), the number of software processes (from 1 to 256) and the number of hardware cores (from 1 to 128) along the strong and weak scaling measures described in this paper.

### Distributed generation of reproducible connections and external "thalamic" stimulus

We mention a feature that has been of some importance to simplify the execution of repeatable strong and weak scaling measures, while varying the number of processes and hardware resources (e.g. processors). We mean, the capability to initialize in a distributed manner an identical network and provide, again in a distributed manner, the same external "thalamic" stimulus to a network composed by a given grid of neural columns, distributed over a varying number of software processes and hardware processors. In a system with N total neurons, distributed among H software processes we can assign a fair share of locN = N/H neurons per software process, and the global and local identities of neurons can be easily computed using the local identifiers of processes and neurons. If there is a grid of CFT = CFX x CFY neural columns, and this info is known to each process, it will be possible for each process to generate forward connectivity patterns that does not depend on the number of processes/hardware processors. The same can be done to generate patterns of external "thalamic" stimulus to the network, e.g. prescribing the number of events per ms per neural column.

### Production of Observables. The behavior of individual neurons.

The DPSNN-STDP code can produce files tracing several observables (list of individual spiking times and spiking neuron identity, mean spiking rates, membrane potentials, synaptic values). The code is equipped with facilities for distributed measurement of time spent on execution of individual routines and sections of the code based on the `MPI_WTIME` function.

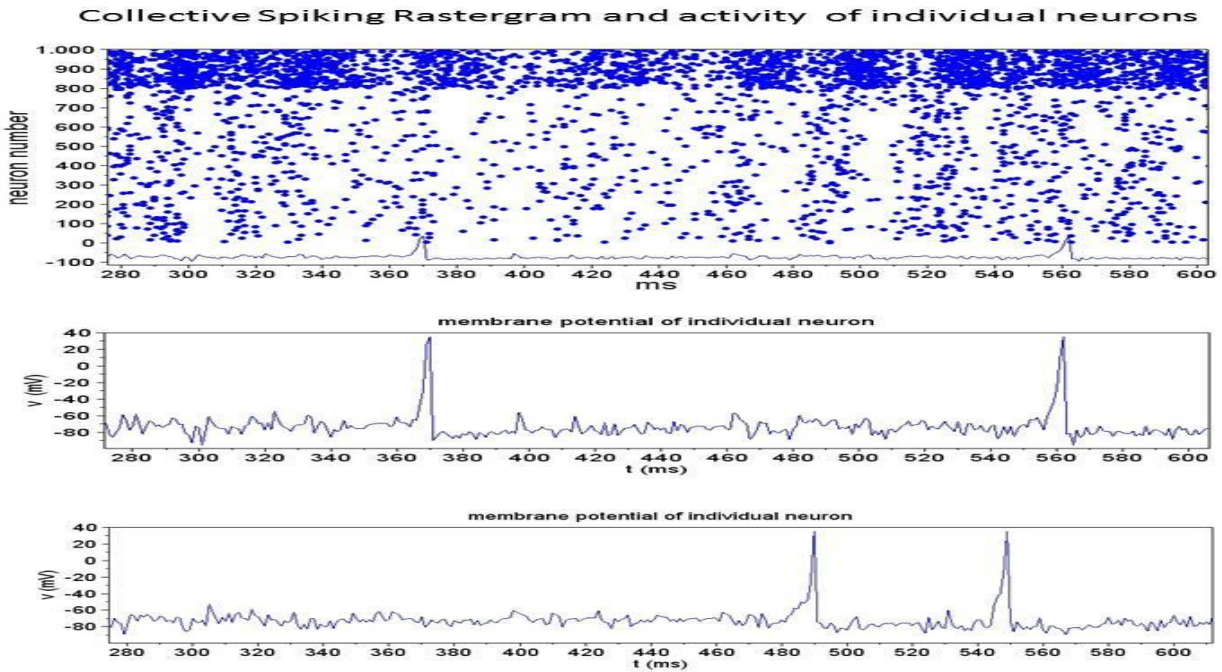

Figure 2-3. A sample trace of 320 ms of spiking activity, produced by the DPSNN-STDP code. In this case, it is the simulation of a single neural column, composed by 1000 neurons (80% excitatory RS, 20% inhibitory FS Izhikevich neurons). Above, each dot in the rastergram represents a spiking event. Below, the traces of the membrane potential of two excitatory neurons.

If necessary, the trace of the evolution of the membrane potential and other state variables of individual neurons can be activated.

The membrane potential of a "resting" neuron fluctuates around a -70mV potential as a result of its own activities and of the perturbation produced by signals produced by other neurons. When the neuron decides to "fire", its membrane potential (v state variable) starts to climb to positive voltages. If a "spike" happened, the Izhikevich rules drops the voltage down to its after spike potential, but the internal variable u keeps a memory of the past. This event is the "spike" of an individual neuron, which is propagated by the axon, and reaches a set of synapses after a time delay, specific for each synapse, which depends on the distance travelled. When reached by the spike, each synapse produces a perturbation of the membrane potential of the target, which depend on its "strength" (here, we consider the simplistic case of the neural soma and dendritic arborization represented by a single "compartment", i.e. a single u(t)-v(t) pair in the case of the Izhikevich equation).

The picture reports about 320 ms of collective spiking activity of a single column of 1000 neurons (800 RS excitatory, 200 FS inhibitory), and the evolution of the individual membrane potential of two neurons. In the "rastergram" the horizontal axis is the simulation time, the vertical axis the identifier of individual neurons. Each dot in the rastergram represents a spiking event.

**Representation of spiking messages**

Spiking messages are sent using an address event representation (AER): we send "axonal spike" messages that carry the identifiers of spiking neurons and are packed in groups that have the same spike emission time and the same target process (i.e. same target cluster of neurons). Our strategy is to defer as much as possible the arborization of the "axon", to reduce the load on the network and unnecessary wait barrier (i.e. waiting for the completion of computations of cluster of neurons from which a process does not expect messages). To this purpose, we perform some preparatory actions during the network initialization phase (performed once at the beginning of the simulation), to reduce the number of active communication channels during the iterative simulation phase.

### Initial construction of the connectivity infrastructure

During the initialization phase, each process can contribute to create the awareness about the subset of processes that should be listened to, during next simulation iterations. At the end of this construction phase, each "target" process should know about the subset of "source" processes that need to communicate with it, and should have created its database of locally incoming axons and synapses. A simple implementation of the construction phase can be realized using two steps.

During the first step, each source process informs other processes about the existence of incoming axons and about the number of incoming synapses to be established. A single word, the synapse counter, is communicated among pairs of processes. Under MPI, this can be achieved by an `MPI_Alltoall()`. Performed once, and with a single word payload, the cost of this first step, creates a cumulative network load proportional to the square of the number of processes. The cost of this operation is negligible in the range of processes explored by this paper.

The second step transfers the identities of synapses to be created on each target process. Under MPI, the payload, a list of synapses specific for each pair in the subset of processes to be connected, can be transferred using a call to the `MPI_alltoallv()` library function. The cumulative load created by this second step is proportional to the product between the total number of processes and the subset of target processes reached by each source process.

The first step produces two effects: 1- it reduces the cost of initial construction of synapses, second step of the construction phase; 2- the knowledge about the non existence of a connection between a pair of processes can be used to reduce the cost of spiking transmission during the simulation iterations.

### Delivery of spiking messages during the simulation phase

Here, we describe the present implementation of the delivery of spiking messages. In this first implementation, we did not take advantage of the possibility of delivering spikes to targets just before the deadline imposed by the synaptic specific delay. Instead, we used a synchronous approach: all spikes are delivered to target processes before proceeding to the simulation of the next time iteration of the neural dynamic.

The delivery of spiking messages can be split in two steps, with communications directed toward subsets of decreasing sizes.

During the first step, single word messages (spike counters) are sent to the subset of potentially connected target processes. On each pair of source-target process subset, the individual spike counter informs about the actual payload (i.e. axonal spikes) that will have to be delivered, or about the absence of spikes to be transmitted between the pair. The knowledge of the subset has been created during the first step of the initialization phase, described in a previous section.

The second step uses the spiking counter info to establish a communication channel only between pairs of processes that actually need to transfer an axonal spikes payload during the current simulation time iteration.

On MPI, both steps can be implemented using calls to the `MPI_Alltoallv()` library function. However the two calls establish actual channels among sets of processes of decreasing size, as described just above.

For the simple bidimensional grid of neural columns and for the mapping on processes used in this experiment this implementation demonstrated to be quite efficient, as reported by the measures presented in the "Results" section, further refined in the "Discussion" section. However, we expect that the delivery of spiking messages will be one of the key point still to be optimized when white area "connectomes" will be introduced, describing the communication channels among a multiplicity of remote cortical areas.

## 3. Results

This section presents: a) the strong and weak scaling of the DPSNN-STDP mini-app benchmark, run on a small scale commodity cluster; b) the profiling of the time-spent on main code components (computational and inter-process communication blocks); c) how the execution time changes depending on the number of synapses projected by each neuron.

We run on a cluster of sixteen dual socket quad core servers, interconnected through a 40 Gb/s commodity network, for a maximum of 128 physical cores running 2.4 GHz.[2]. Each physical core supported two simultaneous threads.

During this experiment, for each neural network size, we checked that the list of spiking neurons and their timings were identical for all run performed using a variable number of software processes and/or physical cores.

In a first set of configurations used for the scaling measures, each neurons projected 200 forward synapses. Neurons were grouped in "columns", each column composed of 1000 neurons (80% excitatory, 20% inhibitory).

In a second set of measures, we also varied the number of synapses projected by each neuron (from 100 to 10000 synapses per neuron) and the number of neurons per column (from 1000 to 12800 neurons per column).

Table 1. We report in the table a subset of the measures used to produce the strong and weak scaling graphs reported in a following section. In particular the table reports: 1- a subset of the configurations and 2- a subset of the measured execution times. We run different problem sizes, from 200 K synapses to 6.6 billion synapses. Each neural network size (a column in the table) was distributed using a varying number of MPI processes, and run on a varying number of physical computational resources. The simulation of a given network size produced an identical spiking and plasticity behavior (e.g. firing activity) over 2000 ms of simulated activity, for all distributions among software processes and/or hardware cores.

| Total synapses | 200 K | 800 K | 3.2 M | 12.8 M | 51.2 M | 204.8 M | 819.2 M | 3.2 G | 6.6 G |
|---|---|---|---|---|---|---|---|---|---|
| Total neurons | 1 K | 4 K | 16 K | 64 K | 256 K | 1024 K | 4.096 M | 16.4 M | 32.8 M |
| Grid of neural columns | 1 x 1 | 2 x 2 | 4 x 4 | 8 x 8 | 16 x16 | 32 x 32 | 64 x 64 | 128x128 | 256x128 |
| Mean firing rate (Hz) | 27 | 24 | 26 | 23 | 22 | 23 | 20 | 22 | 19 |
| Used cores[3] (min-max) | 1-8 | 1-32 | 1-128 | 1-128 | 1-128 | 1-128 | 4-128 | 64-128 | 64-128 |
| MPI processes | 1-8 | 1-32 | 1-128 | 1–256 | 1-256 | 1-256 | 4-256 | 64-256 | 128 |
| Execution time[4] (execution sec / simulated sec) | 0.15 | 0.4 | 1.80 | 3.05 | 6.85 | 20.0 | 59 | 211 | 386 |
| Normalized execution time[5]: execution time / (firing rate × total syn × simulated second) | $2.73 \times 10^{-8}$ | $5.36 \times 10^{-9}$ | $2.41 \times 10^{-8}$ | $4.22 \times 10^{-9}$ | $6.0 \times 10^{-9}$ | $4.22 \times 10^{-9}$ | $3.61 \times 10^{-9}$ | $2.94 \times 10^{-9}$ | $3.07 \times 10^{-9}$ |

As described in a previous section, neural columns were organized in bidimensional grids, and each excitatory neuron projected part of its synapses (76%) toward neurons, and the remainder was distributed among first, second and third neighboring columns, with decreasing proportions (3% toward each of the 4 first neighbouring columns, 2% toward each second neighbour, 1% toward each third neighbour).

---

[2]Each server is a 1U SuperMicro X8DTG-D. Each node in the cluster is a dual socket. Each socket hosts one quad-core Intel(R) Xeon(R) CPU E5620 (max clock @ 2.40GHz). On each core HyperThreading is enabled (two threads per core). Each node is equipped with a Mellanox InfiniBand board, the MT26428 [ConnectX VPI PCIe 2.0 5GT/s - IB QDR (40Gb/s data rate)]. 16 nodes are connected using a Mellanox Switch.
[3] Each cores @2.4 Ghz part of a quad-core Intel(R) Xeon(R) E5620.
[4] Using the "max" number of cores reported in this table
[5] See the "Strong scaling" section of the document for a discussion of the unit of measure for the normalized execution speed.

We varied the size of the bidimensional grid of columns from a minimum configuration of 1x1 to a maximum of 256x128 neural columns. This way, we obtained configurations with a varying number of synapses: from a minimum of 200 K synapses to a maximum of 6.6 G synapses. Each network was distributed on a variable number of software processes, and then assigned to a variable number of physical cores.

Each MPI process hosted a maximum of 1024 neural columns (i.e. 1024 K neurons and 51.2 M synapses per process), and a minimum of 1/8 of neural column (i.e. 125 neurons and 2.5 K synapses per process). Due to the hardware support for two simultaneous threads per core, we observed that, for larger networks (i.e. when more than 16K neurons can be allocated on each process), the best execution time was reached when two MPI software processes were launched on each physical core.

The time step of the simulation was set to 1 ms, but the update of the neural membrane potential was performed using a time step of 0.5 ms. We measured the execution time needed to simulate 2 seconds of spiking activity and synaptic plasticity (i.e. 2000 simulation iterations, using a ms time step). Specifically, we measured the wall clock needed to simulate the $8^{th}$ and $9^{th}$ second of evolution (i.e. we left the system evolve for 8000 simulation steps, before taking our measure). During the 2 seconds selected for the measures, the networks exhibited mean firing rates in the range between 19 and 27 Hz.

The distributed DPSNN-STDP code demonstrated to be competitively fast. For example, only 115 wall clock seconds are required to simulate one second of activity and plasticity of a network composed by 1.6 G synapses (8 M neurons spiking at a mean firing rate of 22.5 Hz) on 128 hardware cores running @ 2.4 GHz. Looking for further improvements and for possible showstoppers when scaling to larger configurations, we profiled the relative weight of the computational and inter-process multicast blocks composing the application, and analyzed the strong and weak scaling behavior, as discussed in the following sections.

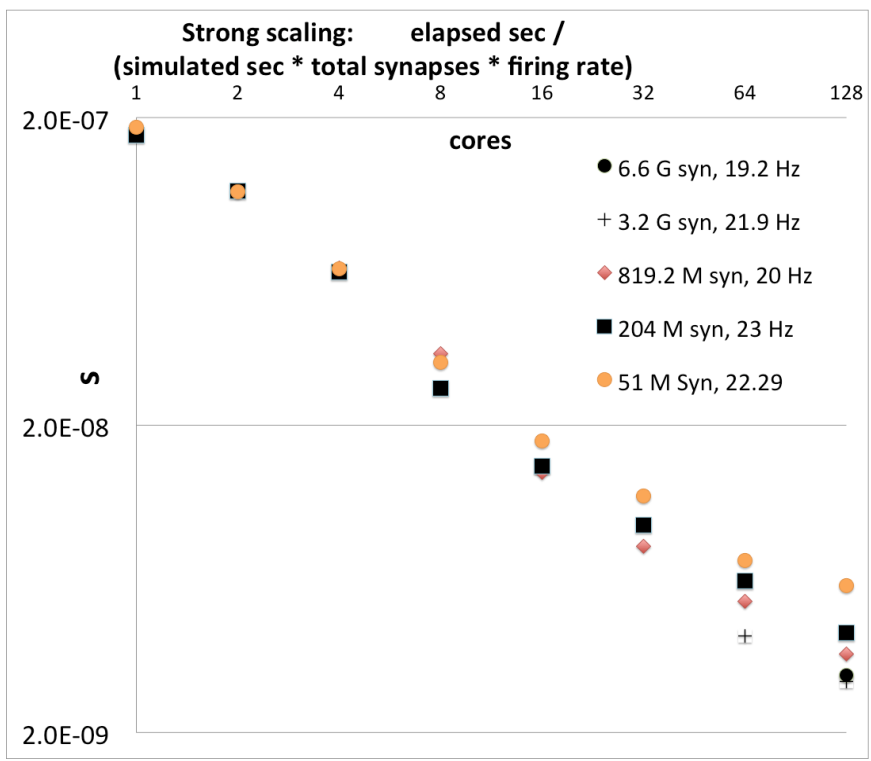

Figure 3-1. Strong scaling of the DPSNN-STDP mini-app benchmark. For an ideal scaling the execution time should grow proportionally to the number of synapses and to the firing rate, and should reduce proportionally to the number of cores applied to the execution.

### Strong scaling

The computational load needed to solve the simulation problem is expected to be proportional: a) to the number of simulated synapses, b) to the firing rate and c) to the physical time to be simulated. Therefore, if we divided the measured execution time on a given number of computational cores by the product of a, b and c, we should obtain an estimate of the execution time needed per synapse per second. In first approximation, we could expect this number to be similar for different problem sizes. Then, an ideal code distributed on an ideal machine should half its execution time when doubling the number of computational cores assigned to the solution of the problem. Here below a picture of the scaling we observed running the DPSNN-STDP code.

However, in Figure 3-1, we observe a scaling on the log-log graph, which, is still not ideal. For example, let us discuss the measures of the 204 M synapse case. On the configuration with 128 hardware cores, 20 seconds were needed to simulate one seconds of spiking and plasticity at a mean firing rate of 23 Hz; while 828 second were needed to produce the same results on a single core. In Figure 3-1, this corresponds to a reduction from $1.75 \times 10^{-7}$ s to $4.22 \times 10^{-9}$ s of the wall clock time needed per synapse, normalized to the firing rate. Instead of decreasing by a factor 128, the actual speed-up is only 41.5. At each doubling in the number of cores, the actual speed-up is only 1.7. Using this slope as representative, the multiplication of the number of cores by a factor 128 would decrease the simulation time 3 times worst that an ideal scaling.

We analyze the results in the "Discussion" section of this document.

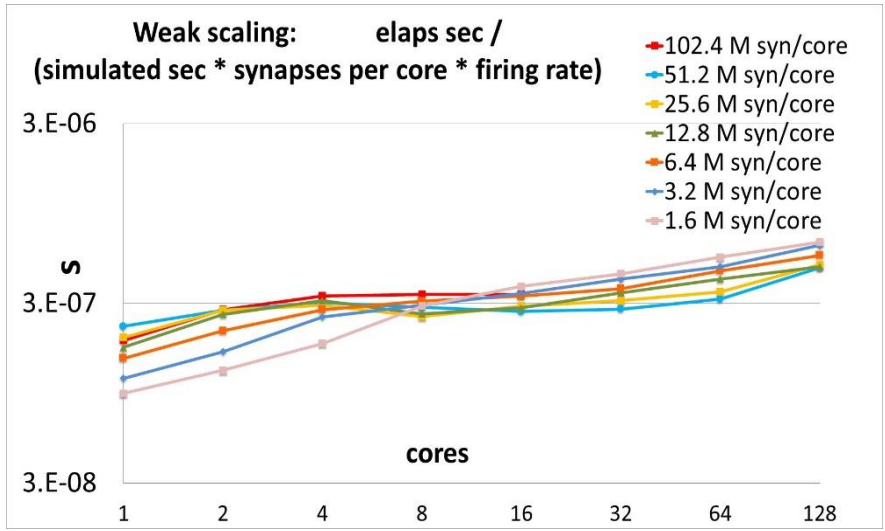

Figure 3-2. Weak scaling of the DPSNN-STDP benchmark.

**Weak scaling**

If, instead of dividing the execution times by the total number of synapses, we divided it by the number of synapses assigned to each computational core, we should obtain, for the same consideration of the previous section, a value that for an ideal code, executing on an ideal machine should be constant for different network sizes and number of computational cores assigned to the solution of the problem. Figure 3-2 is the graph of our measures.

Looking for hints about how to further enhance the scalability features of the system, we observe an interesting feature: when the simulation is distributed among a large number of hardware cores, the simulation runs relatively faster for larger configurations (i.e. configurations that host more "neural columns" per hardware core/software process).

We analyze this point in the next "Discussion" section.

**Profiling of the relative weight of the computational and inter-process multicast blocks**

As discussed in the "Methods" section, each process iterates over a few main computation and inter-process multicast blocks, described in Figure 2-1.

Here, in Table 2, we report the relative time spent on each block during execution, using the same division in functional blocks.

Table 2. Profiling of execution time of the DPSNN-STDP mini-application benchmark: relative weights of computation and communication blocks. The relative weight demonstrated to be quite stable for different network sizes and number of cores used for the simulation. Quite interesting is the weight of the (optional) synchronization barrier. If the barrier is removed, a greater weight would appear in the "Spikes dim" phase. The data used for the table here inserted are relative to a network of 100 M synapses executed by 64 MPI processes on 32 cores[6].

| Function of the block | Relative execution time | Note |
|---|---|---|
| Long term potentiation + after spike dynamic | (9.7 ±0.7)% | Gather[7] + computation |
| Barrier (optional)[8] | (29.9±6.1)% | Workload fluctuations |
| Communication: inter-process multicast: Spikes dim | (0.77±0.10)% | Message passing |
| Communication: inter-process multicast: Spikes payload | (0.82±0.20)% | Message passing |
| Axonal to synaptic spikes: intra-process multicast | (16.8±2.3)% | Dereferencing[9] |
| Add synaptic currents + long term depression | (19.2 ±2.7)% | Computation |
| Thalamic input[10] | 0.01% | Simplified model |
| Ordinary neural dynamic | (11.8±1.4)% | Computation |
| Rastergram & other statistical functions | (1.9±0.1)% | Computation |
| Long term synaptic plasticity | (9.2±1.8)% | Computation |

### Increasing the number of synapses projected by each neuron toward biological figures

In previous sections we verified that the execution time of the DPSNN_STDP simulator is proportional to a) to the total number of simulated synapses, b) to the firing rate and c) to the physical time to be simulated. We also reported about its strong and weak scaling features when the number of processing cores used for the simulation is increased. In Table 1 the number M of synapses projected by each neuron was kept constant (M=200) for all configurations. However, it can be necessary to increase M up to several thousands of synapses per neuron, e.g. when moving toward a detailed biological representation of the human cortex.

Table 3. The number of synapses projected by each neuron has been varied (from M=100 to M=10000). As desired, the execution time per synapse improves when the number of synapses projected by each neuron increases toward biologically realistic figures.

| M: Synapses projected by each neuron | 100 | 200 | 1 000 | 1 000 | 10 000 |
|---|---|---|---|---|---|
| **Relative execution time** | **1.00** | **0.77** | **0.90** | **0.74** | **0.64** |
| Total synapses | 104 M | 104 M | 164 M | 164 M | 128 M |
| Total neurons | 1.04 M | 524 K | 164 K | 164 K | 12.8 K |
| Neurons per cortical module | 1 024 | 1 024 | 10 240 | 1 280 | 12 800 |
| Cortical modules | 1 024 | 512 | 16 | 128 | 1 |
| Grid of cortical modules | 32 x 32 | 32 x 16 | 4 x 4 | 16 x 8 | 1 x 1 |
| Mean firing rate (Hz) | 12.78 | 10.86 | 10.21 | 9.34 | 9.95 |
| Used cores (min-max) | 32 | 32 | 32 | 32 | 32 |

---

[6] See the notes to Table 1 for what concerns the hardware configuration used for the measures.
[7] The benchmarked software implementation is based on sparse accesses from the target neuron to the global list of incoming synapses. In a hardware implementation, based on several independent memory banks, if all synapses incoming to the same neuron were stored in contiguity, this task could be easily accelerated.
[8] If the barrier is not inserted, the time spent waiting for the processes that are still completing their computations appears as a contribution to the time measured for the "spikes dim" communication phase. We verified that the deviations from ideal strong scaling, can be entirely attributed to the cost of measured fluctuations in workload execution (represented by the Barrier block) and to (a very small) cost of communications.
[9] The task to be performed is an "intra-process" multicast, from axons to specific lists of synapses. Instead, the benchmarked software implementation is based on two levels of dereferencing.
[10] In this simulation the thalamic input is computed using a simple statistical model. Actually, this is one of the interface between the neural network and the "external" world, so its weight would greatly increase and add to that of other interfaces to be added.

Table 3 reports the relative execution times for different numbers of projected synapses per neuron (M = {100, 200, 1000 and 10000}).

As desired, the execution time per simulated synapse improves when the number of synapses projected by each neuron increases toward the figure required for simulating the human cortex.

## 4. Discussion

**Impact of unbalances in individual process workload and execution environment**

As discussed in the "Strong scaling" section, a typical behavior of our code is the following: the execution time reduces by a factor 42, when running on 128 cores, compared to the time needed to execute the same configuration on a single core. Initially, we had the temptation to attribute the deviation from ideal scaling to the cost of communications, the usual suspect in the implementation of distributed applications. Therefore, we took great care in optimizing the communication algorithm. After this optimization, the improved communication mechanism described by the "Delivery of spiking messages during the simulation phase" section of this documents produces an expected low cost of communication for the simulated biological connectivity (a bidimensional grid of "cortical modules" with first, second and third neighbor communication). Indeed, the insertion of barriers, when measuring the relative time spent by each process in the set of computation and inter-process multicast blocks, demonstrated that the major cause for deviation from an ideal scaling is the fluctuation in the execution time of different processes, which force some processes to wait for others. Indeed, we verified that the deviations from ideal strong scaling, reported in Figure 3-1, can be entirely attributed to the cost of measured fluctuations in workload (represented by the Barrier block in Table 1) and to (small) cost of communications. There are a few main sources for the difference in execution times, including: 1- the workload of the individual process (i.e. the activity of a cluster of neurons and synapses) changes, depending on the local mean spiking rate and on the number of incoming spikes; 2- the execution time fluctuates even for fixed input and output spikes, due to the software execution environment and to the specificities of the individual hardware node whereon the process is executing. Therefore, we expect load-balancing to play an important role in the simulation of biologically realistic, large scale networks.

**A quantitative basis to drive the development of specialized hardware and software**

During the development and optimization of the code, we used the measure of the relative execution times (like those reported in Table 2) to drive both the definitions of the functional blocks and the time dedicated to their optimization. Hints from frequent measures of the strong and weak scaling behaviour (Figure 3-1 and Figure 3-2) has been also instrumental to the optimization of the code, in particular for the design of the inter-process communication. The data structures of each process and those inside each functional block, as well as the structure adopted for the execution of each block have been designed and coded in a manner that should simplify further distribution/parallelization of the activity of each block. Therefore, we consider the present release of the code (April 2014, svn rev 806) a good starting point for the development of specialized hardware and software architectures.

**From a bidimensional grid of cortical modules to a long-range, complex connectome**

Here, we measured the performance of the DPSNN-STDP mini application benchmark exercised on a bidimensional grid of cortical modules. This interconnection is representative of the local structure, inside a cortical area. However, the simulation of long range "white-matter" connectomes, i.e. of the interconnection among cortical areas, will increase the number of target processes and the complexity of an efficient algorithm for spike delivery.

## 5. Conclusions

We introduced a natively distributed mini-application benchmark representative of plastic spiking neural network simulators. It can be used to measure performances of existing computing platforms and to drive the development of future parallel/distributed computing systems dedicated to the simulation of plastic spiking networks. The mini-application is designed to generate identical spiking behaviors and network topologies over a varying number of processing nodes, simplifying the quantitative study of scalability on commodity and custom architectures. Here, we presented the strong scaling, weak scaling and computation/communication profiling of the DPSNN-STDP code executed on a small scale cluster of commodity processors (varying the number of used physical cores and the problem size). We searched for the cause of the deviation from the ideal scaling behavior for a simple neural network structure (bidimensional grids of neural columns, connected to first, second and third neighboring columns). The cause of the deviation seems to be the jitter in the execution time of each process. This can be due to two main factors: 1) an intrinsic difference between processes (they differ from each other, for example, in the local spiking rate), 2) some extrinsic differences (due to several factor such as the impact of the operating system, the local adaptation of the processor clock speed, etc.). This analysis is valid for the bidimensional topology of neural columns (and software processes). In a future work, we will study the impact of the support of more complex connectomes on the communication, which, in the present release, seems to be well optimized for scalability, at least for the explored size of neural networks and hardware resources.

More in general, our expectation is that the potential performance improvements from dedicated software and hardware co-design solutions will grow, when more complex interconnection topologies (e.g. inter-areal connectomes) will be simulated. The mini-application has been designed to be easily interfaced with standard and custom software and hardware communication interfaces and permit easy measurements of scalability. It has been designed from its foundation to be natively distributed and parallel, and should not pose major obstacles against distribution and parallelization on several platforms. During 2014, we will further enhance it to enable the description of larger networks, more complex connectomes, other neural and synaptic models and prepare it for distribution to a larger community. Meanwhile, the DPSNN-STDP mini-application benchmark will be validated against biologically significant cases.

## 6. Acknowledgements

This work has been partially supported by the EURETILE European integrated FP7 project grant no. 247846. We also acknowledge the contribution of the cooperation with Paolo Del Giudice and Maurizio Mattia, in the framework of the FP7 project CORTICONIC (grant no. 600806).